\documentclass[aps,superscriptaddress,pra,showpacs,twocolumn]{revtex4} %

\usepackage{amsmath,graphicx}

\begin{document}

\title{Tomographic Quantum Cryptography}

\author{Yeong~Cherng~Liang}
\affiliation{Department of Physics, National University of
Singapore, 2 Science Drive 3, Singapore 117542}

\author{Dagomir~Kaszlikowski}
\affiliation{Department of Physics, National University of
Singapore, 2 Science Drive 3, Singapore 117542}

\author{Berthold-Georg~Englert}
\affiliation{Department of Physics, National University of
Singapore, 2 Science Drive 3, Singapore 117542}

\author{Leong~Chuan~Kwek}
\affiliation{National Institute of Education, Nanyang
Technological University, 1 Nanyang Walk, Singapore 639798}
\affiliation{Department of Physics, National University of
Singapore, 2 Science Drive 3, Singapore 117542}

\author{C.~H.~Oh}
\affiliation{Department of Physics, National University of
Singapore, 2 Science Drive 3, Singapore 117542}

\date{5 May 2003}

\begin{abstract} 
We present a protocol for quantum cryptography in which the data obtained for
mismatched bases are used in full for the purpose of quantum state tomography.
Eavesdropping on the quantum channel is seriously impeded by requiring that
the outcome of the tomography is consistent with unbiased noise in the
channel.
We study the incoherent eavesdropping attacks that are still permissible and
establish under which conditions a secure cryptographic key can be
generated.
The whole analysis is carried out for channels that transmit quantum systems
of any finite dimension.
\end{abstract}


\maketitle

\section{Introduction}\label{sec:Intro}
The objective of quantum cryptography is the distribution of a secure
cryptographic key between two parties, traditionally called Alice and Bob.
The key consists of a truly random sequence of ``letters.''
The most important among the schemes proposed for this purpose---the BB84
protocol of Bennett and Brassard \cite{BB84}, and Ekert's E91 protocol 
\cite{E91}---and all experimentally realized schemes 
(Refs.\
\cite{Jennewein+4:00,Naik+4:00,Tittel+3:00,Kurtsiefer+6:02,Waks+6:02},
in particular, but also others), 
use a binary alphabet, i.e., just two letters that are usually denoted 
by the numbers $0$ and $1$.
A very readable account of the state of this art is the recent review article
by Gisin \textit{et al.\/} \cite{Gisin+3:02}.

Binary keys suffice, of course, for all practical purposes and they are
relatively easily generated with the aid of \emph{qubits} (binary
quantum alternatives). 
In fact, the selected experiments cited above provide
increasing evidence that it may be commercially viable to
introduce feasible quantum cryptographic systems in the near
future.

The utter simplicity of the kinematics of a qubit, the most elementary
quantum degree of freedom, facilitates both the theoretical analysis and the
experimental implementations.
And yet, there is a natural curiosity about schemes for quantum cryptography
that exploit richer degrees of freedom, especially \emph{qutrits} (ternary
quantum alternatives, for three-letter keys), and generally \emph{qunits}
($n$-fold quantum alternatives, for $n$-letter keys with $n=2,3,4,\dots$).

Almost all qunit schemes are generalizations of the familiar BB84 and E91 
qubit protocols \cite{Bruss+1:02,Cerf+3:02,Kaszlikowski+6:03,Durt+3:03}, 
and we deal with a particular generalization of E91 in the present paper.
It is worth mentioning, however, that there is also at least one
higher-dimensional scheme of quite a different kind, namely the deterministic
protocol of Beige \textit{et al.\/} \cite{Beige+3:02}, in which 
four-dimensional systems (pairs of qubits, for instance) are used for the
generation of a binary key.

The BB84 and E91 protocols are \emph{in}deterministic because key letters are
only obtained when Alice's and Bob's measurement bases match and,
therefore, a substantial fraction of the data is not used at
all (in BB84) or just for security checks (in E91).
In the protocol we analyze here, all measurement results for mismatched bases
are exploited for complete quantum state tomography, by which 
Alice and Bob manage to impose very stringent conditions on the quantum
channel and so limit eavesdropper Eve's possibilities substantially. 

The paper is organized as follows. 
In Sec.~\ref{sec:protocol} the stage is set by defining the tomographic
protocol. 
Then we analyze, in Sec.~\ref{sec:Eve}, what the eavesdropper can do and
achieve, which prepares the subsequent determination of the security criterion
in Sec.~\ref{sec:CKcriterion}.
We close with a summary of our results and a critical discussion of some
crucial details.

\section{The tomographic protocol}\label{sec:protocol}
\begin{figure}[!t]
\centering\rule{0pt}{4pt}\par
\includegraphics{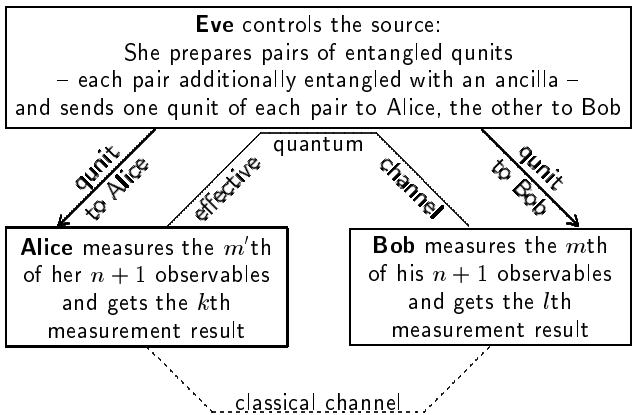}  
\caption{\label{fig:setup}%
Schematic setup of the key distribution system.
Alice and Bob are connected to each other by an effective quantum channel,
which consists of a source that distributes entangled qunit pairs.
For each qunit, Alice measures one of her tomographically complete
observables, chosen at random when the qunit arrives.
Bob does the same for each of his qunits.
They exchange well chosen information about their measurements
through a classical channel, and conclude then whether or not 
the quantum channel has the right characteristics to allow for 
the generation of a secure cryptographic key from their raw data.
In their security analysis, Alice and Bob assume that all imperfections 
of the quantum channel result from eavesdropper Eve's intervention and, 
to be on the safe side, they grant Eve full control of the source.  
}
\end{figure}

We consider a setup of the kind sketched in Fig.~\ref{fig:setup}.
A source emits entangled pairs of qunits to Alice and Bob, 
who receive one qunit each of every pair.
The qunits distributed by the source in this manner constitute an effective 
quantum channel between Alice and Bob (\AB), although these two users are not 
themselves sending any quantum systems to each other.
As a consequence of unavoidable imperfections, both in the functioning of the
source and in the transmission line, this quantum channel will be noisy to
some extent, so that \AB\ will not receive qunit pairs with the
ideal properties they hope for. 

Nevertheless, they will be able to generate a secure cryptographic key if the
noise level is below a certain threshold.
But to be on the safe side, they must determine this threshold level 
under the assumption that all imperfections result from their adversary Eve's 
intervention, who eavesdrops on the communication between \AB.
In particular, one must grant Eve full control over the qunit-pair source, 
and she will try to know as much about the qunits detected by \AB,
as the laws of physics allow her to know.  

After receiving a qunit from the source, Alice measures a non-degenerate 
observable that she selects at random from her set of $n+1$ tomographically 
complete observables~\cite{fn:n+1a,fn:n+1b}.
She keeps a private record of the observables she measures  
and of the outcomes of her measurements.
Likewise, Bob measures on each of his qunits an observable randomly chosen 
from his corresponding set, and keeps a record of his data as well.
We adopt the notation of \cite{MKp} and
denote by $\ket{m_k}$ the $k$th eigenket of Alice's $m$th observable
and by $\ket{\overline{m}_k}$  the $k$th eigenket of Bob's $m$th observable.
The correspondence between the two sets of observables, or rather between the
orthonormal measurement bases they provide, is then established by
requiring that  
\begin{equation}\label{eq:correspond}
\braket{0_j}{m_k}=\braket{\overline{m}_k}{\overline{0}_j}
\end{equation}
holds for $j,k=0,1,2,\dots,n-1$ and $m=0,1,2,\dots,n$.
In short, the roles of bras and kets are interchanged.

Ideally, \AB\ wish to receive from the source 
the maximally entangled two-qunit state $\ket{\psi}$ that is specified by
\begin{align}\label{eq:psidef}
\ket{\psi}&=\frac{1}{\sqrt{n}}\sum_{k=0}^{n-1}\ket{0_k\overline{0}_k}
=\frac{1}{\sqrt{n}}\sum_{k=0}^{n-1}\ket{1_k\overline{1}_k}=\cdots\nonumber\\
&=\frac{1}{\sqrt{n}}\sum_{k=0}^{n-1}\ket{n_k\overline{n}_k}\,.
\end{align}
As a consequence of (\ref{eq:correspond}), 
it has the same form regardless of the pair of observables that
is used to define it.

When the transmission is over, \AB\ announce their choice
of observables, their respective $m$ values, for all qunits through 
a public channel. 
They can then divide the detected qunit pairs into two
groups, one in which the measurement bases match 
(both $m$ values are the same, which happens with a
probability of $1/(n+1)$ ), and one in which the bases do not match. 
In the absence of noise, the measurement results of the first group
(the respective $k$ values---referred to as \emph{nit values}) are
perfectly correlated and thus give rise to a cryptographic key in
an alphabet with $n$ letters.

In reality, however, \AB\ must take into account Eve's attempts at
eavesdropping and the resulting disturbance of the quantum channel.
As a consequence thereof, the statistical properties of the 
detected qunit pairs will not be correctly described by the pure two-qunit
state of (\ref{eq:psidef}). 
Rather than the projector $\ketbra{\psi}$, an appropriate statistical
operator $\rho$ applies to the qunit pairs emitted by a non-ideal 
source.

Since \AB\ measure tomographically complete sets of observables
on their respective qunits, they can determine the actual two-qunit state
$\rho$ from their measurement results. 
They exploit all data of the mismatched bases for this purpose, 
and some of the matched-bases data.
Ideally, they wish for the projector $\ketbra{\psi}$ 
but, realistically, they expect to find a $\rho$ of the form 
\begin{equation} \label{eq:source}
\rho=(\beta_0-\beta_1)\ketbra{\psi}+\frac{\beta_1}{n}
\quad\mbox{with}\enskip\beta_0+(n-1)\beta_1=1\,,
\end{equation}
which is what one gets when an imperfect transmission line admixes 
unbiased noise to $\ketbra{\psi}$.
The non-negative parameters $\beta_0$ and $\beta_1$ have the following 
physical significance:
$\beta_0$ is the probability that Alice and Bob get the same nit value 
when the bases match, and $\beta_1$ is the probability that
Bob gets a particular one of the $n-1$ values that are different from 
Alice's nit value.

Formally, $\rho$ is a non-negative operator of unit trace, and thus
permissible as a statistical operator, 
whenever ${0\leq\beta_1/\beta_0\leq n/(n-1)}$.
But only values in the range
\begin{equation}
  \label{eq:BetaRange}
  0\leq\frac{\beta_1}{\beta_0}\leq1
\end{equation}
correspond to an admixture of symmetric noise to $\ketbra{\psi}$ and,
therefore, this is the parameter range of interest.
The limiting values mark the extreme situations of ``no noise at all''
($\beta_0=1$, $\beta_1=0$) and ``nothing but noise'' ($\beta_0=\beta_1=1/n$).

Sources that emit two-qunit states of a kind different from (\ref{eq:source}) 
are not regarded as trustworthy by Alice and Bob.
As the crucial, defining step of the \emph{tomographic protocol},
they thus accept the raw key sequence only if their state tomography
confirms that the source emits a two-qunit state of the form (\ref{eq:source}).
Otherwise, they reject the data wholly and use a different source.

\section{Eavesdropping}\label{sec:Eve}
\subsection{Choosing the right ancilla states}\label{sec:ancilla}
By imposing this rather stringent requirement, \AB\ restrict Eve's 
possibilities markedly. 
Her strategy is to keep a quantum record of what she sends
to \AB\ by entangling each qunit pair with an ancilla, and to
perform a judiciously chosen measurement on the ancilla after carefully
weighing the information exchanged by \AB\ through the public channel. 
Quite generally, Eve's option is to prepare an entangled pure state of the form
\begin{equation}\label{eq:mPsi1}
\ket{\Psi}=\sum_{k,l=0}^{n-1}\ket{m_k\overline{m}_l}
\ket{\tilde{\mathrm{E}}^{(m)}_{kl}}
\qquad\mbox{(any $m=0,1,\dots, n$)},  
\end{equation}
where the $\ket{\tilde{\mathrm{E}}_{kl}^{(m)}}$'s are
the unnormalized kets of the ancilla states attached by Eve
(with reference to the $m$th pair of \AB's observables). 
Since there is no advantage in generating a mixed state instead, it is
sufficient to consider all such pure-state preparations.

Now, the two-qunit state received by \AB\ is obtained by 
tracing $\ketbra{\Psi}$ over the ancilla degree of freedom, 
and their insistence on getting $\rho$ of (\ref{eq:source}) 
implies that Eve must choose her ancilla states such that they obey
\begin{align}
\braket{\tilde{\mathrm{E}}^{(m)}_{kl}}{\tilde{\mathrm{E}}^{(m)}_{k'l'}}
&=\frac{\beta_0-\beta_1}{n}\delta_{kl}\delta_{k'l'}
+\frac{\beta_1}{n}\delta_{kk'}\delta_{ll'}
\notag\\[1ex]
&=\left\{
\begin{array}{cl}
\beta_0/n & \mbox{if $k=l=k'=l'$,}\\
\beta_1/n & \mbox{if $k=k'\neq l=l'$,}\\
(\beta_0-\beta_1)/n & \mbox{if $k=l\neq k'=l'$,} \\
0 & \mbox{otherwise.}
\end{array}
\right.\label{eq:innerEtilde}
\end{align}
The right-hand side does not depend on the $m$ value to which the ancilla
states on the left refer and, therefore, the mapping 
$\ket{\tilde{\mathrm{E}}^{(m)}_{kl}}\to\ket{\tilde{\mathrm{E}}^{(m')}_{kl}}$
is unitary.
Quite explicitly, two different sets of ancilla states are related to each
other by  
\begin{equation}
  \label{eq:2ancSets}
    \ket{\tilde{\mathrm{E}}^{(m)}_{kl}}=
\sum_{k',l'}\ket{\tilde{\mathrm{E}}^{(m')}_{k'l'}}
\braket{m_k}{m'_{k'}}\braket{m'_{l'}}{m_l}\,,
\end{equation}
which is an immediate consequence of (\ref{eq:mPsi1}) 
and (\ref{eq:correspond}).
As a check of consistency, one can exploit the completeness and
orthonormality of the single-qunit states $\ket{m_k}$ to  
verify rather easily that (\ref{eq:innerEtilde}) holds for any $m$ value, 
if it holds for one of them.
In summary, then, it does not matter which $m$ value Eve chooses in 
(\ref{eq:mPsi1}).

It is expedient to introduce normalized ancilla states 
$\ket{\mathrm{E}^{(m)}_{kl}}$ in
accordance with
\begin{eqnarray}\label{eq:AncStates}
k=l:&&\ket{\tilde{\mathrm{E}}_{kk}^{(m)}}
=\ket{\mathrm{E}^{(m)}_{kk}}\sqrt{\beta_0/n}\,,
\nonumber\\    
k\neq l:&&
\ket{\tilde{\mathrm{E}}_{kl}^{(m)}}
=\ket{\mathrm{E}^{(m)}_{kl}}\sqrt{\beta_1/n}\,.    
\end{eqnarray}
Then
\begin{equation}\label{eq:NormAncStates}
\braket{\mathrm{E}^{(m)}_{kl}}{\mathrm{E}^{(m)}_{k'l'}}
=\left\{
\begin{array}{cl}
1& \mbox{if $k=k'$ and $l=l'$,}\\
1-\beta_1/\beta_0& \mbox{if $k=l\neq k'=l'$,} \\
0 & \mbox{otherwise,}
\end{array}
\right.
\end{equation}
and
\begin{align}
\ket{\Psi}=&\sqrt{\frac{\beta_0}{n}}
\sum_{k=0}^{n-1}\ket{m_k\overline{m}_k}\ket{\mathrm{E}^{(m)}_{kk}}
\notag\\&+\sqrt{\frac{\beta_1}{n}}\sum_{k\neq l}
\ket{m_k\overline{m}_l}\ket{\mathrm{E}^{(m)}_{kl}}
\quad\mbox{(any $m=0,\dots, n$).}
\label{eq:mPsi2}
\end{align}
As stated in (\ref{eq:NormAncStates}), for each $m$ value, 
the ancilla states $\ket{\mathrm{E}^{(m)}_{kl}}$ with $k\neq l$ 
are orthogonal to each other and orthogonal to the ones with $k=l$.
The latter are not orthogonal among themselves
(except when $\beta_1=\beta_0$, the case of pure noise 
and of very little interest), but rather have the same inner products for all
pairs,
\begin{equation}
  \label{eq:pyra1}
k\neq l:\quad\braket{\mathrm{E}^{(m)}_{kk}}{\mathrm{E}^{(m)}_{ll}}
=1-\frac{\beta_1}{\beta_0}\,.  
\end{equation}
The $n$ ancilla states $\ket{\mathrm{E}^{(m)}_{kk}}$ are thus 
linearly independent, except when $\beta_1=0$, which is the ideal 
situation of no noise at all, that is no eavesdropping \cite{fn:2ndException}.

\subsection{Ancilla subspaces}\label{sec:ancilllaSpace}
This exception aside, the $k=l$ ancilla states $\ket{\mathrm{E}^{(m)}_{kk}}$
span a $n$-dimensional subspace that is orthogonal to the
$(n^2-n)$-dimensional subspace spanned by the ${k\neq l}$ states.
We refer to them as the \emph{first} and the \emph{second} subspace, 
respectively.
The subspaces associated with different $m$ values are related to
each other by the unitary transformations of (\ref{eq:2ancSets}).

Eve takes advantage of the structure of these subspaces in the eavesdropping
attack that we now proceed to describe.
We shall deal solely with attacks, in which she performs measurements 
on the ancillas one-by-one, commonly termed \emph{incoherent} attacks.
By contrast, in a \emph{coherent} attack, she would measure some joint
observables of a few, or perhaps many, ancillas \cite{fn:CorrPairs}.
This limitation is mainly dictated by the technical difficulties that one
faces when analyzing coherent attacks. 
We note, however, that some have argued---notably 
Cirac and Gisin \cite{Cirac+1:97}, and Wang \cite{Wang:02}---that
coherent attacks are not more powerful than incoherent attacks, but their
arguments refer rather explicitly to protocols of the BB84 type, 
with intercept-resend eavesdropping attacks, and are not 
immediately applicable to our tomographic qunit protocol.   
 
Eve's incoherent eavesdropping procedure is as follows.
The information exchanged by \AB\ over the classical channel identifies
those qunit pairs that contribute to the raw key sequence, 
the ones for which Alice's $m$ value is the same as Bob's.
To find out, as much as she can, about the nit values that \AB\ have recorded
for each of these matched qunit pairs, Eve performs a suitably chosen
measurement on the respective ancillas, one at a time.
The statistical operator for one of these ancillas is obtained by tracing
$\ketbra{\Psi}$ over the qunit degrees of freedom, with the outcome
\begin{equation}
  \label{eq:rhoEve0}
  \rho^{(m)}_\mathrm{Eve}=
\frac{\beta_0}{n}\sum_{k=0}^{n-1}\ketbra{\mathrm{E}^{(m)}_{kk}}
+\frac{\beta_1}{n}\sum_{k\ne l}
\ket{\mathrm{E}^{(m)}_{kl}}\bra{\mathrm{E}^{(m)}_{kl}}\,,
\end{equation}
where $m$ identifies the matched pair of bases.
The first summation corresponds to the situation, in which \AB\ get the same
nit value and the ancilla ends up in the first subspace, which happens with
probability $\beta_0$. 
And the situation of differing nit values, when the final ancilla state 
is in the second subspace, is accounted for by the second summation, which
carries the complementary weight of $(n-1)\beta_1=1-\beta_0$.

Since the various $\rho^{(m)}_\mathrm{Eve}$'s ($m=0,1,\dots,n$) are unitarily
equivalent, it is sufficiently general to consider just one $m$ value. 
For notational simplicity, we leave it implicit from here on 
and suppress the $m$ label.
Then we have
\begin{equation}
  \label{eq:rhoEve}
\rho^{\ }_\mathrm{Eve}=\beta_0\rho^{(=)}
+(1-\beta_0)\rho^{(\neq)}
\end{equation}
with
\begin{equation}
  \label{eq:rhoEve1}
  \rho^{(=)}=\frac{1}{n}\sum_{k=0}^{n-1}\ketbra{\mathrm{E}^{\ }_{kk}}
\end{equation}
and
\begin{equation}
  \label{eq:rhoEve2}
\rho^{(\neq)}=\frac{1}{n(n-1)}\sum_{k\ne l}
\ket{\mathrm{E}^{\ }_{kl}}\bra{\mathrm{E}^{\ }_{kl}}\,.
\end{equation}
The first of these conditional statistical operators, $\rho^{(=)}$, 
applies when \AB\ have the same nit value and the second, $\rho^{(\neq)}$, 
applies when they don't. 
Since $\rho^{(=)}$ and $\rho^{(\neq)}$ reside in the first and second 
subspace, respectively, Eve can discriminate between the two situations 
unambiguously.

Suppose she thus establishes that different nit values are the case. 
Under this circumstance, 
she performs a measurement that distinguishes between the $k\neq l$ 
ancilla states, which is surely possible because they are mutually orthogonal.
She finds the ancilla in the state with ket $\ket{\mathrm{E}_{kl}}$, say,
and then knows with certainty that Alice's nit value is $k$, and Bob's is
$l$ (with $k\neq l$, of course).

By contrast, if Eve establishes that the nit values of \AB\ are the same, 
she cannot find out with certainty what is this common value because 
the $k=l$ ancilla states are not orthogonal to each other, 
except when the $\beta_1=\beta_0$ limit is reached in (\ref{eq:BetaRange}) 
and the right-hand side vanishes in (\ref{eq:pyra1}).

For $\beta_1<\beta_0$, 
Eve's attempts in discerning the non-orthogonal $\ket{\mathrm{E}_{kk}}$ 
ancilla states in the first subspace are prone to error. 
Recalling (\ref{eq:pyra1}), we note that the inner product for each pair 
of them is the same positive number, just like it is for 
the vectors pointing from the tip of a pyramid to the corners at its base. 
It is known that the error-minimizing measurement for such ``pyramid states''
is the so-called \emph{square-root measurement} (see, e.g., 
\cite{Chefles:00,Barnett:01} and the pertinent references therein, 
in particular \cite{Helstrom:76} and \cite{Holevo:78}).
Although demonstrating this optimality requires a careful argument, 
it is easy to grasp the basic idea of a square-root measurement.

\subsection{Square-root measurement}\label{sec:SRM}
For the following, up to and including (\ref{eq:SRM12}), 
we restrict the discussion to the first subspace. 
Then the $n$ kets
\begin{equation}
  \label{eq:SRM1}
  \ket{\mathrm{e}_{kk}}=\frac{1}{\sqrt{n\rho^{(=)}}}\ket{\mathrm{E}_{kk}}
\end{equation}
decompose the identity by construction, 
\begin{equation}
  \label{eq:SRM2}
  \sum_k \ketbra{\mathrm{e}_{kk}}=1,
\end{equation}
and thus define a generalized measurement---the square-root measurement.
Now note the eigenvalue equations
\begin{subequations}\label{eq:SMR3}
\begin{eqnarray}
\left(\rho^{(=)}-r_0\right)\sum_j\ket{\mathrm{E}_{jj}}&=&0\,,
\label{eq:SRM3a} \\  \label{eq:SRM3b}
\left(\rho^{(=)}-r_1\right)\left(\ket{\mathrm{E}_{kk}}
-\frac{1}{n}\sum_j\ket{\mathrm{E}_{jj}}\right)
&=&0
\end{eqnarray}
\end{subequations}
with 
\begin{equation}
  r_0=1-\frac{n-1}{n}\frac{\beta_1}{\beta_0}\,,
\qquad
r_1=\frac{\beta_1}{n\beta_0}\,,
  \label{eq:SRM4}
\end{equation}
so that
\begin{equation}
  \label{eq:SRM5}
   r_0+(n-1)r_1=1\,,\qquad r_0-r_1=1-\frac{\beta_1}{\beta_0}\,.
\end{equation}
The eigenvalue $r_0$ is nondegenerate, 
whereas $r_1$ is ${(n-1)}$-fold, and not $n$-fold, because the 
$n$ kets in (\ref{eq:SRM3b}) have a vanishing sum.
We make use of these eigenvalues in writing
\begin{equation}
  \label{eq:SRM6}
  \frac{1}{\sqrt{\rho^{(=)}}}=
\frac{r_0+\sqrt{r_0r_1}+r_1-\rho^{(=)}}{\sqrt{r_0r_1}(\sqrt{r_0}+\sqrt{r_1})}
\end{equation}
and then exploit (\ref{eq:SMR3}) to establish
\begin{equation}
  \label{eq:SRM7}
 \ket{\mathrm{e}_{kk}}=\frac{1}{\sqrt{nr_1}}\left(\ket{\mathrm{E}_{kk}}
-\frac{1-\sqrt{r_1/r_0}}{n}\sum_j\ket{\mathrm{E}_{jj}}\right)\,.
\end{equation}

The parameters $\eta_0$ and $\eta_1$ that appear in the probability amplitudes
\begin{equation}
  \label{eq:SRM8}
  \braket{\mathrm{e}_{kk}}{\mathrm{E}_{ll}}
=\sqrt{\eta_0}\,\delta_{kl}+\sqrt{\eta_1}(1-\delta_{kl})
\end{equation}
are crucial, inasmuch as they quantify Eve's knowledge about the common nit
value of \AB: Upon finding $\ket{\mathrm{e}_{kk}}$, she knows that the actual
nit value is $k$ with probability $\eta_0$ and that it is either one of the
${n-1}$ other values with probability $\eta_1$.
In conjunction with (\ref{eq:SRM5}), the required normalization
\begin{equation}
  \label{eq:SRM9}
  \eta_0+(n-1)\eta_1=1
\end{equation}
follows immediately from the explicit expressions
\begin{equation}
  \label{eq:SRM10}
  \sqrt{\eta_0}=\frac{\sqrt{r_0}+(n-1)\sqrt{r_1}}{\sqrt{n}}\,,\qquad
\sqrt{\eta_1}=\frac{\sqrt{r_0}-\sqrt{r_1}}{\sqrt{n}}\,.
\end{equation}
The implied identity
\begin{equation}
  \label{eq:SRM11} 
   \sqrt{\eta_0}-\sqrt{\eta_1}=\sqrt{\frac{\beta_1}{\beta_0}}
\end{equation}
is worth noting.

We close the discussion of the square-root measurement in the first subspace
with the observation that Eve's reference states $\ket{\mathrm{e}_{kk}}$ are 
orthonormal,
\begin{equation}
  \label{eq:SRM12}
  \braket{\mathrm{e}_{kk}}{\mathrm{e}_{ll}}=\delta_{kl}\,.
\end{equation}
As a consequence, the generalized measurement defined by the decomposition
(\ref{eq:SRM2}) is in fact a standard von Neumann measurement. 

Now returning to the general discussion of the full statistical operator
(\ref{eq:rhoEve}), we summarize Eve's strategy as follows.
She performs a measurement that distinguishes the $n^2$ states~\cite{fn:remind}
\begin{equation}
  \label{eq:AllEveStates}
  \ket{\mathrm{e}_{kl}}=\left\{\begin{array}{l}
  \text{as given in (\ref{eq:SRM7}) for $k=l$,}\\[1ex]
  \ket{\mathrm{E}_{kl}} \mbox{\ for $k\neq l$,}
    \end{array}\right.
\end{equation}
which are orthonormal.
With probability $(n-1)\beta_1=1-\beta_0$ she finds a state with $k\neq l$, 
and then infers that Alice's nit value is $k$, and Bob's is $l$.
And when Eve finds a $k=l$ state, which happens with probability $\beta_0$,
she knows that \AB\ have the same nit value and can guess it right 
with probability $\eta_0$, but will guess a particular one 
of the ${n-1}$ wrong values with probability $\eta_1$.

\subsection{Probabilities}\label{sec:Probs}
In more formal terms, the joint probability that, for matched bases, Alice
gets nit value $k$, Bob gets $l$, and Eve detects $\ket{\mathrm{e}_{k'l'}}$
is given by
\begin{eqnarray}
  \label{eq:JointProbs}
  p^{\ }_{kl;k'l'}=
\biglb|\braket{\widetilde{\mathrm{E}}_{kl}}{\mathrm{e}_{k'l'}}\bigrb|^2
&=&\frac{\beta_0}{n}\delta_{kl}\delta_{k'l'}
\bigl[(\eta_0-\eta_1)\delta_{kk'}+\eta_1\bigr]
\nonumber\\
&&
+\frac{\beta_1}{n}(1-\delta_{kl})\delta_{kk'}\delta_{ll'}\,.
\end{eqnarray}
All reduced and conditional probabilities are derived from this expression 
by partial summation and normalization. 
For later reference we note the joint probabilities for Alice and Bob,
\begin{equation}
  \label{eq:JointProbs-AB}
    p^{\mathrm{(A\&B)}}_{kl}=\sum_{k',l'}  p^{\ }_{kl;k'l'}=
\frac{1}{n}\bigl[(\beta_0-\beta_1)\delta_{kl}+\beta_1\bigr]\,,
\end{equation}
and for Alice and Eve,
\begin{eqnarray}
  \label{eq:JointProbs-AE}
    p^{\mathrm{(A\&E)}}_{k;k'l'}=\sum_l  p^{\ }_{kl;k'l'}&=&
\frac{\beta_0}{n}\delta_{k'l'}\bigl[(\eta_0-\eta_1)\delta_{kk'}+\eta_1\bigr]
\nonumber\\
&&+\frac{\beta_1}{n}\delta_{kk'}(1-\delta_{k'l'})\,,
\end{eqnarray}
as well as the individual probabilities for Alice and Bob,
\begin{equation}
  \label{eq:Probs-A+B}
   p^{\mathrm{(A)}}_{k}=\sum_{l,k',l'}  p^{\ }_{kl;k'l'}=\frac{1}{n}\,,\quad
   p^{\mathrm{(B)}}_{l}=\sum_{k,k',l'}  p^{\ }_{kl;k'l'}=\frac{1}{n}\,,
\end{equation}
and for Eve,
\begin{equation}
  \label{eq:Probs-E}
   p^{\mathrm{(E)}}_{k'l'}=\sum_{k,l}  p^{\ }_{kl;k'l'}=
\frac{1}{n}\bigl[(\beta_0-\beta_1)\delta_{k'l'}+\beta_1\bigr]\,.
\end{equation}

To get a first rough understanding of the significance of \AB's probabilities
${\beta_0,\beta_1}$ and Eve's conditional probabilities ${\eta_0,\eta_1}$,
consider this scenario.
A qunit pair has been received by \AB\ and detected with matched bases.
Both Bob and Eve are asked to bet on Alice's nit value. 
Bob's best strategy is to guess that Alice's value agrees with his own, and he
guesses right with probability $\beta_0$, but he is never sure about Alice's
nit value.
Eve, by contrast, knows Alice's nit value with certainty when detecting the
ancilla in one of the $k\neq l$ states of (\ref{eq:AllEveStates}), 
and guesses right with probability $\eta_0$ otherwise. 
Her total betting odds are thus ${1-\beta_0+\beta_0\eta_0}$.
The comparison with Bob's establishes that, if such bets are
performed frequently,
\begin{equation}
  \label{eq:betting}
  \begin{array}{l}
  \text{Bob wins more often if $\beta_0>(n+3)\beta_1$,}\\
  \text{Eve wins more often if $\beta_0<(n+3)\beta_1$,}\\
  \text{and they come out even if $\beta_0=(n+3)\beta_1$.} 
  \end{array}
\end{equation}
These betting odds are, however, really only a rough measure of Bob's and
Eve's knowledge about Alice's nit value, because Eve's information is
qualitatively different from Bob's.
As discussed in the next section, the ratio $\beta_1/\beta_0$ must be
substantially below the $1/(n+3)$ threshold of (\ref{eq:betting}) if \AB\
want to be able to generate a secure key from the raw key sequence that these
bets are about.

\section{Security criterion}\label{sec:CKcriterion}
\subsection{Csisz\'ar-K\"orner threshold}\label{sec:CKthresh}
\begin{figure}[!t]
\centering\rule{0pt}{4pt}\par
\includegraphics{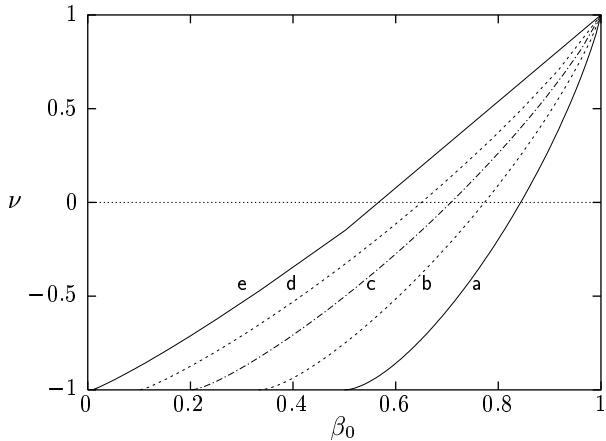}   
\caption{\label{fig:CKyield}%
The difference $\nu$, defined in (\ref{eq:CKyield1}), 
of the mutual information between Alice and Bob and between Eve 
and either one of them, as a function of $\beta_0$, for various values of $n$.
Curves \textsf{a}--\textsf{e} are for $n=2,3,5,10$, and $100$, respectively.
A secure key can be generated from the raw key sequence if $\nu$ is positive.
The threshold value of $\beta_0$, the point of intersection with
the $\nu=0$ line, decreases with increasing $n$ and approaches
$\beta_0=\frac{1}{2}$ for $n\to\infty$.
}
\end{figure}

A more systematic quantitative measure of what Bob and Eve know about Alice's
nit values is the mutual information between the respective parties.
With the probabilities of (\ref{eq:JointProbs-AB}) and (\ref{eq:Probs-A+B}),
we get
\begin{eqnarray}
  \label{eq:MutInf-AB}
  I(\mathrm{A\&B})&=&\sum_{k,l}p^{\mathrm{(A\&B)}}_{kl}
\log_n\frac{p^{\mathrm{(A\&B)}}_{kl}}{p^{\mathrm{(A)}}_{k}p^{\mathrm{(B)}}_{l}}
\nonumber\\
&=&1+\beta_0\log_n\beta_0+(1-\beta_0)\log_n\beta_1
\end{eqnarray}
for the mutual information between Alice and Bob,
where, fitting to the $n$-letter alphabet, 
the logarithm is taken to base $n$.
Likewise, the mutual information between Alice (or Bob) and Eve is given by
\begin{eqnarray}
  \label{eq:MutInf-AE}
  I(\mathrm{A\&E})&=&\sum_{k,k',l'}p^{\mathrm{(A\&E)}}_{k;k'l'}
\log_n\frac{p^{\mathrm{(A\&E)}}_{k;k'l'}}
{p^{\mathrm{(A)}}_{k}p^{\mathrm{(E)}}_{k'l'}}
\nonumber\\
&=&1+\beta_0\bigl[\eta_0\log_n\eta_0+(1-\eta_0)\log_n\eta_1\bigr]\,.
\end{eqnarray}
Their difference
\begin{eqnarray}
  \label{eq:CKyield1}
  \nu&\equiv& I(\mathrm{A\&B})- I(\mathrm{A\&E})
\nonumber\\
     &=&\beta_0\log_n\beta_0+(1-\beta_0)\log_n\beta_1
\nonumber\\
&&      -\beta_0\bigl[\eta_0\log_n\eta_0+(1-\eta_0)\log_n\eta_1\bigr]
\end{eqnarray}
is shown in Fig.~\ref{fig:CKyield} for $n=2,3,5,10$, and $100$ 
over the $\beta_0$ range of (\ref{eq:BetaRange}). 
There, it is a monotonically increasing function of $\beta_0$ that grows 
from $\nu=-1$ for $\beta_0=1/n$ to $\nu=1$ for $\beta_0=1$.  
The values of $\beta_0$, where the sign of
$\nu$ changes, are listed in Table~\ref{tbl:CKthreshold} for some $n$, 
along with the corresponding values of $n\beta_1/\beta_0$ 
and the ratio $\eta_1/\eta_0$ of Eve's conditional probabilities.

\begin{table}[!t]
\caption{\label{tbl:CKthreshold}%
Threshold values of some parameters.
For the various $n$ values of the first column, the table reports values of
 $\beta_0$, $n\beta_1/\beta_0$,  
and $\eta_1/\eta_0$ for which the CK threshold is reached ($\nu=0$), 
or for which the CK yield is 50\% ($\nu=\frac{1}{2}$).
The limiting values for $n\to\infty$ are shown in the last row.}
\begin{ruledtabular}
\begin{tabular}{rcccccc}
& \multicolumn{3}{c}{\hrulefill\,$\nu=0$\,\hrulefill} &
 \multicolumn{3}{c}{\hrulefill\,$\nu=0.5$\,\hrulefill} 
\\
$n$ & $\beta_0$ & $n\beta_1/\beta_0$ & $\eta_1/\eta_0$ 
    & $\beta_0$ & $n\beta_1/\beta_0$ & $\eta_1/\eta_0$ 
\\ 
\hline
  2 & 0.8436 & 0.3707 & 0.2659 & 0.9357 & 0.1373 & 0.4661 \\
  3 & 0.7733 & 0.4398 & 0.2741 & 0.9050 & 0.1574 & 0.4649 \\
  4 & 0.7334 & 0.4846 & 0.2790 & 0.8870 & 0.1698 & 0.4641 \\[1ex]
  5 & 0.7077 & 0.5163 & 0.2821 & 0.8750 & 0.1785 & 0.4635 \\
 10 & 0.6503 & 0.5975 & 0.2880 & 0.8468 & 0.2010 & 0.4604 \\
 30 & 0.6016 & 0.6851 & 0.2887 & 0.8203 & 0.2266 & 0.4532 \\[1ex]
 50 & 0.5881 & 0.7146 & 0.2872 & 0.8123 & 0.2358 & 0.4496 \\
100 & 0.5747 & 0.7475 & 0.2843 & 0.8040 & 0.2462 & 0.4448 \\
$\infty$ & 0.5\phantom{000}& 1 & 
 0.25\phantom{00}  & 0.75\phantom{00} & 0.3333 & 0.4019 
\end{tabular}
\end{ruledtabular}
\end{table}

Now, according to the Csisz\'ar-K\"orner (CK) Theorem~\cite{CK}, 
a secure cryptographic key can be generated from the raw key sequence,
by means of a suitably chosen error correcting code and classical (one-way)
communication between Alice and Bob, if the mutual information between Alice 
and Bob exceeds that between Eve and either of them.
In the present context, this is to say that the tomographic protocol is
secure (under the incoherent eavesdropping attacks considered) if $\nu>0$.
Moreover, $\nu$ is then the yield of the key generation, in the sense that a
secure key of length $\nu L$ can be obtained from a raw key sequence 
of length $L$. 
This invites to call $\max\bigl\{0,\nu\bigr\}$ the \emph{CK yield}.
It is positive when $\beta_0$ is larger than the threshold values of
Table~\ref{tbl:CKthreshold} and vanishes at and below the threshold.

Any actual implementation of the tomographic protocol for quantum key
distribution needs a reasonable efficiency. 
The $\nu=0$ threshold is then of less interest than, say, the $\nu=\frac{1}{2}$
threshold at which the CK yield reaches 50\%. 
The respective values of  $\beta_0$, $n\beta_1/\beta_0$, and $\eta_1/\eta_0$
are also listed in Table~\ref{tbl:CKthreshold}.  

\begin{figure}[!t]
\centering\rule{0pt}{4pt}\par
\includegraphics{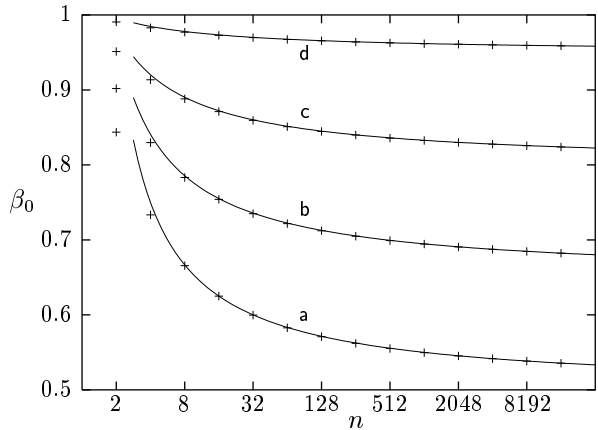}   
\caption{\label{fig:CKthreshold}%
Threshold values of $\beta_0$ for CK yields of 0\%, 30\%, 60\%, and 90\%.
As a function of $n$, with the abscissa linear in $\log n$, the crosses 
display the exact values of $\beta_0$ for which $\nu=0$ (set~\textsf{a}), 
$\nu=0.3$ (set~\textsf{b}), $\nu=0.6$ (set~\textsf{c}), or $\nu=0.9$
(set~\textsf{d}), respectively.
The solid lines show the corresponding values of the analytical approximation
(\ref{eq:beta0Appr}), which is assuredly good for large $n$ values, but
performs remarkably well even for small ones.
}
\end{figure}

For sufficiently large $n$, the threshold values of $\beta_0$ are well
approximated by
\begin{equation}
  \label{eq:beta0Appr}
  \beta_0\approx\frac{1+\nu+\log_n\frac{2}{1-\nu}}
{2+\log_n\frac{1+\nu}{1-\nu}}\,,
\end{equation}
which becomes the strikingly simple $\beta_0\approx\frac{1}{2}(1+\log_n2)$ 
for $\nu=0$.
By comparing with the entries in the second and fifth columns of 
Table~\ref{tbl:CKthreshold}, we observe that the error is 1\% or less 
for $\nu=0$ and $n>4$, or $\nu=\frac{1}{2}$ and $n>3$.
For $\nu=0$, $0.3$, $0.6$, and $0.9$, we illustrate 
(\ref{eq:beta0Appr}) in Fig.~\ref{fig:CKthreshold}.

\subsection{Channel capacities}\label{sec:InfoTh}
It is interesting to view the CK security criterion also
from another perspective of information theory.
Rather than mutual information, the relevant notion 
is then that of channel capacity.

The generation of the raw key can be regarded as the outcome of a 
communication between Alice and Bob through the effective quantum channel 
of Fig.~\ref{fig:setup}. 
By choosing her observables and measuring them, 
Alice effectively prepares the qunits sent to Bob in the states resulting
from the formal procedure of state reduction. 
For instance, after Alice has measured her $m$th observable and found
$\ket{m_k}$ for her qunit, her reduced statistical operator for Bob's qunit 
is 
\begin{equation}
  \label{eq:rhoBob}
  \rho_k^{(\mathrm{B},m)}=(\beta_0-\beta_1)\ketbra{\barr{m}_k}+\beta_1\,,
\end{equation}
with each $k$ value occurring---or, now, \emph{being sent}---with 
probability $1/n$.
Upon measuring his $m$th observable, Bob gets the nit value $k$ with
probability $\beta_0$ and each of the $n-1$ other ones with probability 
$\beta_1$, quite consistent, of course, with the joint probabilities
(\ref{eq:JointProbs-AB}). 

These projective measurements carried out by Bob can be interpreted as his 
attempt to extract the information encoded by Alice in  the states 
$\rho_k^{(\mathrm{B},m)}$, so that, for every $m$, a certain quantum 
channel is thus defined between Alice and Bob. 
Since, for a given $m$, Bob gets all right nit values with the same 
probability $\beta_0/n$, and all wrong values with the same probability 
$\beta_1/n$, we are in fact dealing with a so-called 
\emph{weakly symmetric channel} \cite{Cover+Thomas:91}. 
For a channel of this kind, transmission at full capacity is achieved 
for totally random input, as is the case here. 

All $m$ values are equivalent, and the capacity of each channel, 
$C(\mathrm{A \& B})=1+\beta_0\log_n\beta_0+(1-\beta_0)\log_n\beta_1$, 
is just equal to the mutual information  $I(\mathrm{A\&B})$ of
(\ref{eq:MutInf-AB}) \cite{fn:channels}. 
A similar reasoning applies to the effective ancilla channel between 
Alice and Eve that is associated with Eve's square root measurement. 
The capacity $C(\mathrm{A\&E)}$ of this channel is also equal to
the corresponding mutual information, $I(\mathrm{A\&E)}$ of
(\ref{eq:MutInf-AE}). 

It follows that the CK threshold criterion for the tomographic protocol 
has a simple intuitive meaning:
secure one-way communication is possible if the capacity of the channel 
between Alice and Bob is higher than the capacity of the channel between 
Alice and Eve.

\section{Summary and discussion}
The protocol for quantum key distribution that is described and analyzed in
this paper differs from other protocols by the element of complete quantum
state tomography. 
For this purpose, Alice and Bob exploit the measurement results they obtain
for unmatched bases, rather than just discarding these data as one does in
the BB84 protocol and its various generalizations.
The check for a violation of Bell's inequality in the E91 protocol amounts to
a partial state tomography and, in this sense, our tomographic protocol might
be viewed as a refinement and generalization of the E91 protocol.

In the tomographic protocol, Alice and Bob insist on the source emitting 
entangled two-qunit states of a particular form---only states from a 
one-parametric family are in fact regarded as acceptable---and thereby they 
limit Eve's choice of eavesdropping attacks stringently.
Up to unitary equivalence, there is then only one preparation by Eve 
of the qunit pairs, entangled with her ancilla states, that gives her 
best knowledge of the raw key sequence obtained by Alice and Bob.
But even with this optimized eavesdropping attack, Eve does not acquire
enough information to prevent Alice and Bob from generating a secure key,
provided that the two-qunit state is in the parameter range where the   
Csisz\'ar-K\"orner theorem applies.
Alice and Bob find out whether this is the case when they determine the
parameters of the two-qunit state by state tomography.

But the story does not end here.
If the source emits states outside the parameter regime where an
immediate key generation is possible, Alice and Bob might still be able to
achieve their objective although it seems that Eve knows too much.
They just need to first ``distill'' a better raw key, for which
purpose they can choose between the quantum procedure 
of \emph{entanglement distillation} \cite{Deutsch+5:96,M+PHor:99}
and the classical procedure of \emph{advantage distillation} \cite{Maurer:93}.
Recent work establishes \cite{Bruss+5:03} that both procedures are
applicable if ${\beta_0>2\beta_1}$ and only then, which is, therefore, 
the true threshold condition for the tomographic protocol. 

The square-root measurement, on which the present analysis of Eve's
incoherent attack is based (and also the analysis in \cite{Bruss+5:03}),
maximizes Eve's odds of guessing Alice's nit values right but, as noted by
Shor \cite{Shor:02}, it does not always maximize her information about them.
In other words, it may happen that a (slightly) larger value of
the mutual information between Alice and Eve obtains for another measurement.
The only case on record for which this is known to occur is, however, a
very flat ${n=3}$ pyramid of states, outside the physical parameter range of
(\ref{eq:BetaRange}). 
Other cases are likely to exist, possibly also for larger $n$ values and
rather tall pyramids.
If so, the CK threshold values would be changed (slightly), but presently 
there is no indication that the ${\beta_0>2\beta_1}$ condition for successful
distillation is affected.
These matters are not settled as yet, systematic investigations are being
performed, and results will be reported in due course. 

In protocols of the BB84 type, Alice prepares qunits and sends them to Bob,
with Eve eavesdropping on the quantum channel.
As discussed in Sec.~\ref{sec:InfoTh}, 
one method of preparation could be to detect one qunit of an entangled pair,
thereby reducing the state of the other, which is on its way to Bob.
In the setup of Fig.~\ref{fig:setup} this would amount to having, so to say,
the source inside Alice's laboratory.
It follows that our analysis has a bearing also on schemes of the BB84 type.
Reversing the argument, no matter how Alice prepares the qunit sent to Bob,
she can treat her record of it as if it were the result of a measurement on
another qunit, be it real or virtual.
Alice and Bob can then treat their joint records as if the data referred to
entangled qunit pairs, and apply the tomographic protocol.
In effect, this limits Eve's choice of eavesdropping attacks on the quantum
channel in an analogous way and, as a consequence, our results are also
applicable to tomographic protocols of this other kind.

In the security analysis of protocols of BB84 type, Eve is assumed to
intercept the qunits in transmission, to use some cloning device for copying 
the qunit state with the fidelity permitted by quantum limitations, and to
perform eventually  a suitable measurement on the quantum copy.
It is in this context of single-qunit protocols that relations equivalent to 
(\ref{eq:SRM10}) were first derived for qutrits ($n=3$) by Bru\ss\ and
Macchiavello \cite{Bruss+1:02,fn:convert}, and conjectured to hold for
arbitrary $n$ \cite{Bruss+1:02,Acin+2:03}.
Also, the $\beta_0>2\beta_1$ threshold condition for both distillation
procedures applies to BB84-type qunit protocols  
\cite{Gisin+1:99,Acin+2:03}.
And for the question about the optimality of Eve's square-root measurement 
in the tomographic protocol, there is an analogous question about the optimal 
cloning device in single-qunit protocols.
In view of these close interrelations, a definite answer to one of them 
will surely teach us a lesson about the other question, too.

\begin{acknowledgments}
We gratefully acknowledge valuable discussions with D.~Bru\ss, K.~Chang,
M.~Christandl, T.~Durt, A.~Ekert, A.~Gopinathan, D.~Gosal, 
and C.~Macchiavello.
This work was supported by A$^*$Star Grant No.~012-104-0040.
\end{acknowledgments}


\begin{thebibliography}{99}
\bibitem{fn:n+1a}
The statistical operator of a single qunit has $n^2-1=(n+1)(n-1)$ 
independent real parameters. 
Repeated measurements of one nondegenerate qunit observable establish the
values of $n-1$ parameters and, therefore, any tomographically complete set
of single-qunit observables contains at least a total 
number of $n+1$ independent observables.
Alice and Bob could each use more than this minimum number of observables,
but for the sake of simplicity in notation we assume that they don't. 
Note that a qunit pair would require only $n^2+1$ two-qunit observables for
complete state tomography, rather than the $(n+1)^2$ pairs of single-qunit 
observables that Alice and Bob are using.
\bibitem{fn:2ndException}
Another exception occurs for $\beta_1/\beta_0=n/(n-1)$, but this ratio is
outside the range (\ref{eq:BetaRange}) of physical interest.
\bibitem{fn:CorrPairs}
More generally, one could consider coherent attacks in which Eve prepares
entangled multi-qunit-pair states rather than the single-qunit-pair state 
of (\ref{eq:source}). 
Alice and Bob would then notice correlations between different qunit pairs.
We take for granted that they protect themselves by also looking for 
such correlations at the time when they exchange information for the primary
purpose of state tomography.
\bibitem{fn:remind}
Remember that the $m$ label is suppressed. There is a set of $n^2$ states of
the kind (\ref{eq:AllEveStates}) for each value of $m$.
\bibitem{fn:convert} 
Put $n\to d$, $\beta_0\to1-D$, and $\eta_0\to f_d(D)$ to convert our 
notational conventions to those of \cite{Bruss+1:02}.
\end{thebibliography}
\end{document}